\documentstyle[epsfig]{mn2e}
\begin{document}
\input BoxedEPS.tex
\newcommand{\aip}{{\small ${\cal AIPS}$}}
\newcommand{\gtsim}{\mbox{{\raisebox{-0.4ex}{$\stackrel{>}{{\scriptstyle\sim}}
$}}}}
\newcommand{\ltsim}{\mbox{{\raisebox{-0.4ex}{$\stackrel{<}{{\scriptstyle\sim}}
$}}}}
\newcommand{\s}{$\stackrel{\rm s}{.}$}
\newcommand{\h}{$^{\rm h}$}
\newcommand{\m}{$^{\rm m}$}
\newcommand{\pp}{$\stackrel{\prime\prime}{.}$}
\newcommand{\de}{$^{\circ}$}
\newcommand{\p}{$^{\prime}$}
\newcommand{\arc}{$^{\prime\prime}$}
\newcommand{\marc}{^{\prime\prime}}
\newcommand{\rs}{{\em $r_s$}}
\newcommand{\DPM}{{\em DPM}}
\newcommand{\alf}{{\displaystyle\biggl({\nu_{\rm h} \over \nu_{\rm l}}\biggr)^{\alpha}} }

\newcommand{\figstart}[1]
    { \begin{figure}[htb]
      \begin{picture}(0,#1) }
\newcommand{\figend}[4]
    { \end{picture}
      \special{#1}
      \caption[#2]{#3}
      \label{#4}
      \end{figure} }
\newcommand{\fig}[5]
    { \figstart{#1}
      \figend{#2}{#3}{#4}{#5} }
\newcommand{\bHS}{\beta_{\mbox{\scriptsize HS}}}
\newcommand{\bBF}{\beta_{\mbox{\scriptsize BF}}}
\newcommand{\nT}{\nu_{\mbox{\scriptsize T}}}
\newcommand{\et}{E_{\mbox{\scriptsize T}}}
\newcommand{\nTn}{\nu_{\mbox{\scriptsize Tn}}}
\newcommand{\nTf}{\nu_{\mbox{\scriptsize Tf}}}
\newcommand{\tn}{\tau_{x\mbox{\scriptsize n}}}
\newcommand{\tf}{\tau_{x\mbox{\scriptsize f}}}
\newcommand{\xn}{x_{\mbox{\scriptsize n}}}
\newcommand{\xf}{x_{\mbox{\scriptsize f}}}
\newcommand{\yn}{y_{\mbox{\scriptsize n}}}
\newcommand{\yf}{y_{\mbox{\scriptsize f}}}
\newcommand{\lln}{l_{\mbox{\scriptsize n}}}
\newcommand{\llf}{l_{\mbox{\scriptsize f}}}
\newcommand{\Dn}{f(\Delta_{\mbox{\scriptsize n}})}
\newcommand{\Df}{f(\Delta_{\mbox{\scriptsize f}})}
\newcommand{\B}{\mbox{$B$}}
\newcommand{\Bo}{\mbox{$B$}_{0}}

\SetEPSFDirectory{/scratch/sbgs/figures/hst/}
\SetRokickiEPSFSpecial
\HideDisplacementBoxes

\title[Modelling JWST mid-infrared counts]{Modelling JWST mid-infrared counts II: Extension to 5.6 $\mu$m, optical, radio and X-rays. }
\author[Rowan-Robinson M.]{Michael Rowan-Robinson\\
Astrophysics Group, Blackett Laboratory, Imperial College of Science 
Technology and Medicine, Prince Consort Road,\\ 
London SW7 2AZ}
\maketitle
\begin{abstract}
In Paper I (Rowan-Robinson 2024), models derived in 2009 to fit mid-infrared (8-24 micron) source counts from the IRAS, ISO and Spitzer missions, 
were found to provide an excellent fit to deep counts at 7.7-21 $\mu$m with JWST, demonstrating that the evolution of dusty star-forming galaxies 
is well understood. Here the treatment of optical spectral energy distributions (SEDs) is improved and the counts are extended to 5.6 $\mu$m and
optical wavelengths.  The models proved a good fit to the latest, deeper, JWST counts.  The models are also extended to radio and X-ray wavelengths.
Predicted redshift distributions are given for a range of wavelengths and flux-densities.

\end{abstract}


\section{Introduction}

Source-counts at radio, X-ray, infrared and submillimetre wavelengths, combined with the spectrum of the 
infrared background, have given us important constraints on the evolution of AGN (active galactic nuclei) 
and on star-formation history of the universe.  
In an earlier paper (Rowan-Robinson 2024, Paper I), I showed that models derived to fit
{\it IRAS}, {\it ISO}, {\it Spitzer}, {\it Herschel}, and ground-based submillimetre counts 
(Rowan-Robinson 2009) give a good fit to the first 7.7-21 $\mu$m counts from JWST (Wu et al 2023).  
Since then deeper JWST counts have been published by Yang et al (2023), Stone et al (2024), 
Sajkov et al 2024, and Ostlin et al (2025), including counts at 5.6 $\mu$m.  Windhorst et al (2023)
presented JWST counts at 0.85-4.4 $\mu$m.   Paper I gives a review 
of earlier work modelling infrared source-counts.

In this paper I show how the model fits the new data.  By improving the treatment of the template 
spectral energy distributions (SEDs) at optical wavelengths, I extend the range of the models to 
5.6 $\mu$m and to the B-band (0.44 $\mu$m).  I also show fits to radio and X-ray source-counts.

A cosmological model with k=0, $\Lambda$ = 0.7, $H_0$ =72, has been used throughout.

\section{Methodology}

The models are as described in Paper I, except in the treatment of the optical SEDs, where the optical 
templates of Babbedge et al (2004), Rowan-Robinson et al (2008), based on detailed star-formation 
histories for each Hubble type, are used here (Figure 1).  In this process it became clear that what was previously 
treated as a single 'cirrus' or 'quiescent' component needed to be separated into two components: 
post-starburst galaxies, where there has been a starburst within the past $10^9$ years or so, and genuinely 
quiescent galaxies, where there has been no significant star-formation for $> 10^9$ years.  Efstathiou 
and Rowan-Robinson (2003) modelled the optical and mid-infrared SEDs in detail of a selection of both these types 
of galaxy which had SCUBA 850 $\mu$m detections, with a range of star-formation histories, 
optical extinctions $A_V$, and the ratio of the radiation field intensity to that in the solar neighbourhood,
 $\psi$.  For the higher redshift, post-starburst galaxies the range of $A_V$ was 0.2-3.1 and the range of 
$\psi$ was 1-21, while for the more local quiescent galaxies the range of $A_V$ was 0.4-0.9 and the range 
of $\psi$ was 2-8.  Selection at 850 $\mu$m will have biased $A_V$ towards higher values.
As representative of post-starburst galaxies I adopt a starburst SED with $A_V = 0.5$, and as representative 
of quiescent galaxies, I adopt an Sbc SED with $A_V=0$.  These choices were governed by the fits to the counts
and to the integrated background spectrum (Section 6, below).  For both types I adopt $\psi=5$, the value used 
previously for cirrus galaxies, which fits the infrared spectrum of our Galaxy well (Rowan-Robinson 1992).
I also tried $\psi=10$ for post-sb galaxies, but this made the count fits worse. A more sophisticated treatment 
would involve a range of $A_V$ and $\psi$.

The parameters for the M82 and
Arp 220 starbursts were based on detailed fits to the SEDs of the prototype galaxies (Ade et al 2011), except that
the amplitude of the optical component of the M82 SED was reduced by a factor 2 in the light of comparisons of
the 5.6 $\mu$m redshift distribution with JWST data (see section 5 below).

The luminosity functions and evolution rates are as in Paper I except for the treatment of 'cirrus' 
galaxies.  The luminosity functions for each population were derived from the 60 $\mu$m luminosity function via a mixture table,
with is a function of 60 $\mu$m luminosity, as described in Rowan-Robinson (2001).  Each population was allowed its own luminosity 
evolution rate (of the form eqn (1) below) and the parameters $a_0$, P, Q,  were tuned to provide acceptable fits to the source-counts from 8-1100 $\mu$m.
The SEDs for the components were derived from radiative transfer model fits to the {\it IRAS}, {\it ISO} and {\it Spitzer} infrared galaxy
populations. Any galaxy population not present in the {\it IRAS} 12-100 $\mu$m data would not be included in the count predictions. 
 
Here, for post-starburst galaxies, I assume:
\newline

$\phi(z) = [ a_0 + (1-a_0) exp  Q (1-y )]  (y^P -(y_f)^P)$                 (1)
\newline

\noindent
where $y = (t/t_0)$, $y_f = t(z_f)/t_0$, $a_0$ = 0.8, Q = 12, P = 4 , $z_f$ = 10, (all as in Paper I)

and for quiescent galaxies I assume:
\newline

$\phi(z) = 0.6 a_0 (y^P -(y_f)^P)$, (2) 
\newline

where $a_0$ = 0.8, P = 4, $z_f$ = 10.

The luminosity evolution rates for the different components are shown in Fig 2 as a function of cosmic 
time.  The five infrared populations are now:

\begin{itemize}
\item {{\bf starbursts due to major mergers}, with Arp220 as the prototype}
\item {{\bf starbursts due to interactions and minor mergers}, with M82 as the prototype}
\item {{\bf AGN dust tori}}
\item {{\bf post-starburst galaxies}}
\item {{\bf quiescent galaxies} (including ellipticals).}
\end{itemize}

The starbursts dominate at $L_{ir}> 10^{11} L_{\odot}$, and include the extreme starburst studied by Rowan-Robinson et al (2018), 
for which $L_{ir}> 10^{13} L_{\odot}$.

\begin{figure}
\epsfig{file=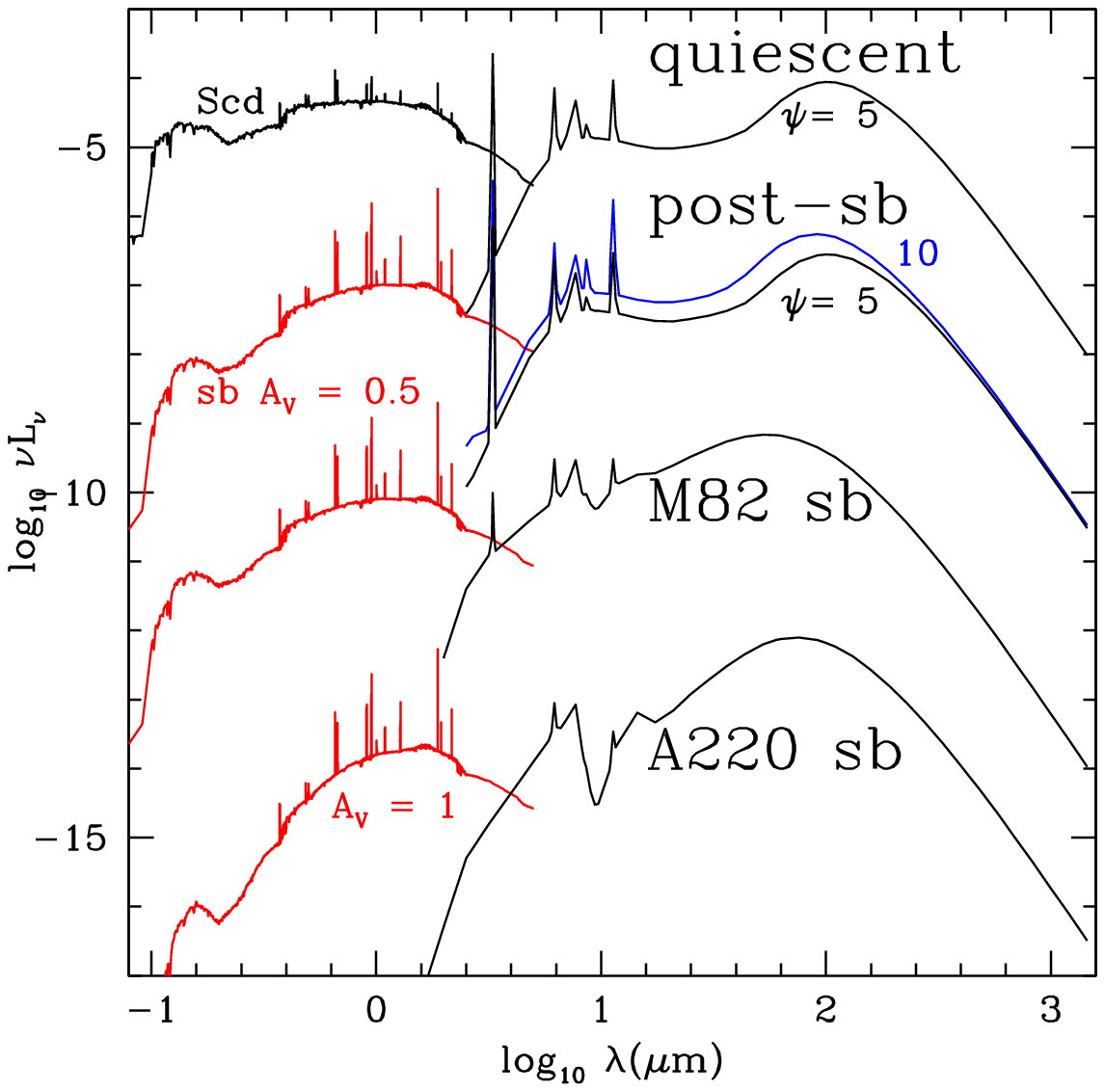,angle=0,width=7.5cm}
\caption{Revised optical seds for cirrus and starburst populations. From the top: quiescent (Scd for optical-nir, radiation field intensity $\psi$ = 5 for mid-ir);  post-sb  (starburst with $A_V$ = 0.5 for opt-air, radiation field intensity $\psi$ = 5 (black, adopted) and 10 (blue) for mid-ir); M82 starburst;
and Arp 220 starburst.
Vertical scaling is arbitrary.
}  
\end{figure}

\begin{figure}
\epsfig{file=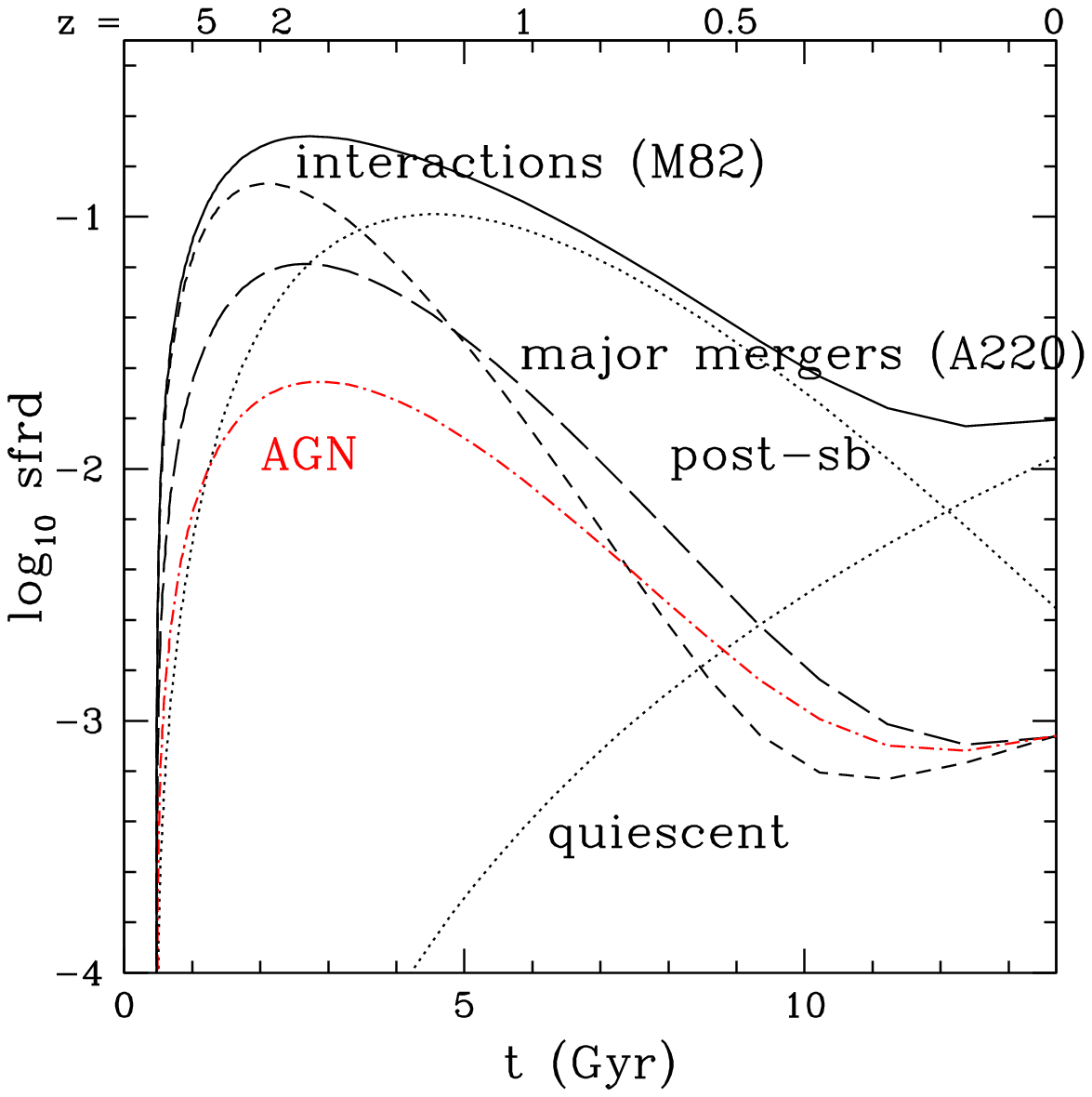,angle=0,width=7.5cm}
\caption{Star-formation density as a function of time, showing interactions and mergers (M82, short-dashed), major mergers (A220, long dashed), post-starburst and quiescent (dotted) phases. Solid locus is total star-formation density, which was compared with observations in Paper I.  The somewhat flatter form of the evolution of the AGN dust torus component is shown by the red 
dash-dotted locus.
}
\end{figure}

\begin{figure*}
\epsfig{file=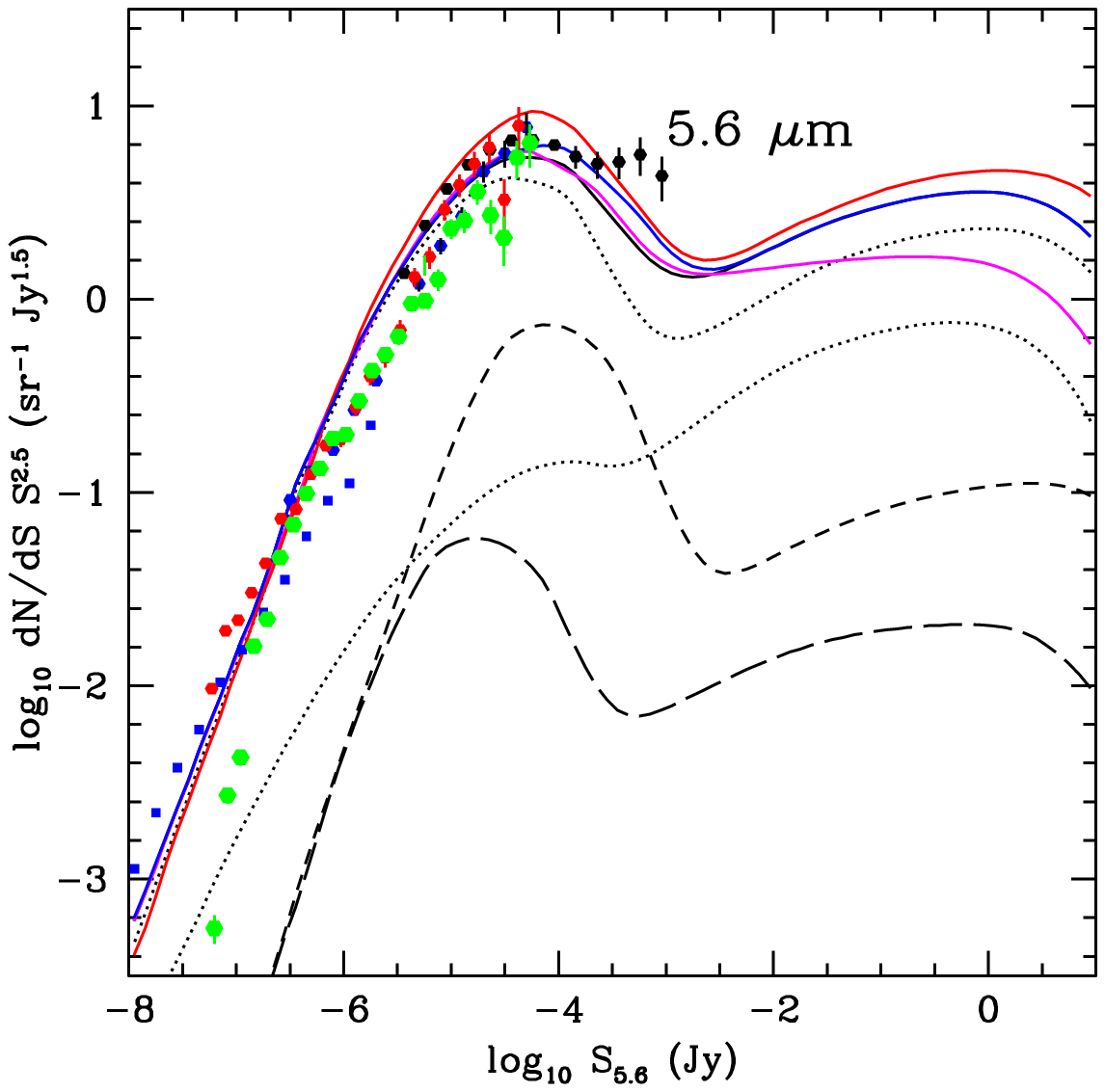,angle=0,width=7.5cm}
\epsfig{file=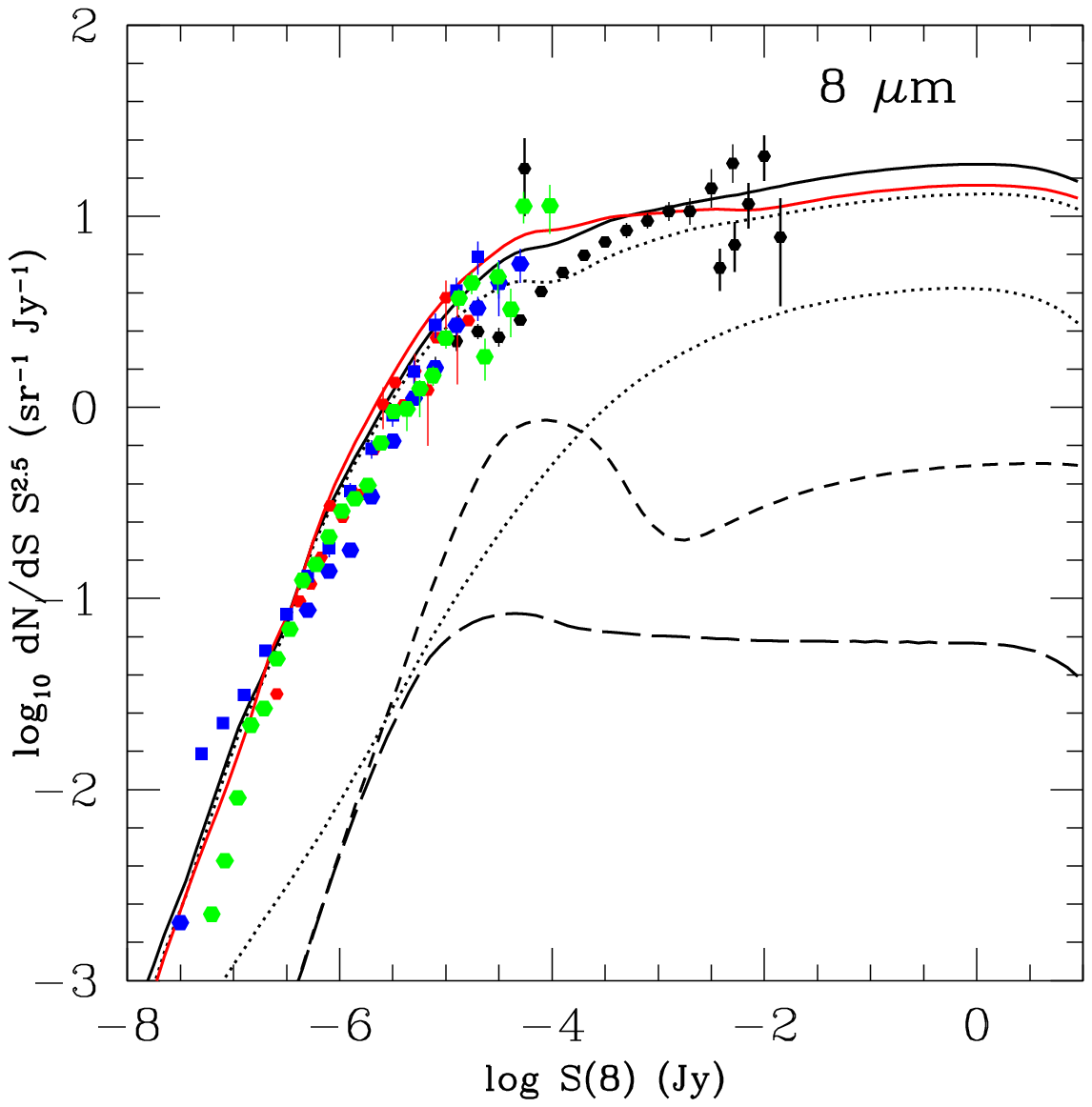,angle=0,width=7.5cm}
\epsfig{file=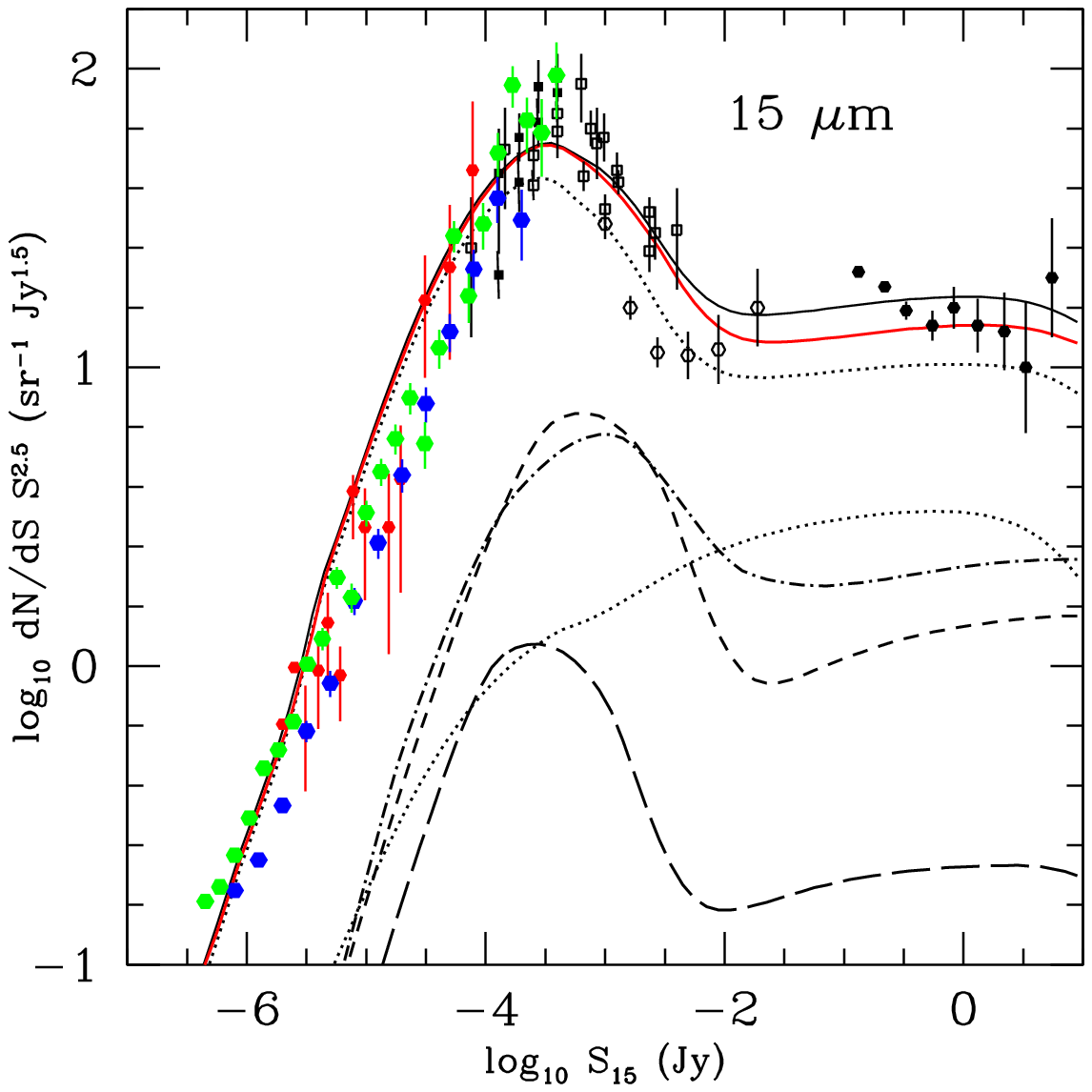,angle=0,width=7.5cm}
\epsfig{file=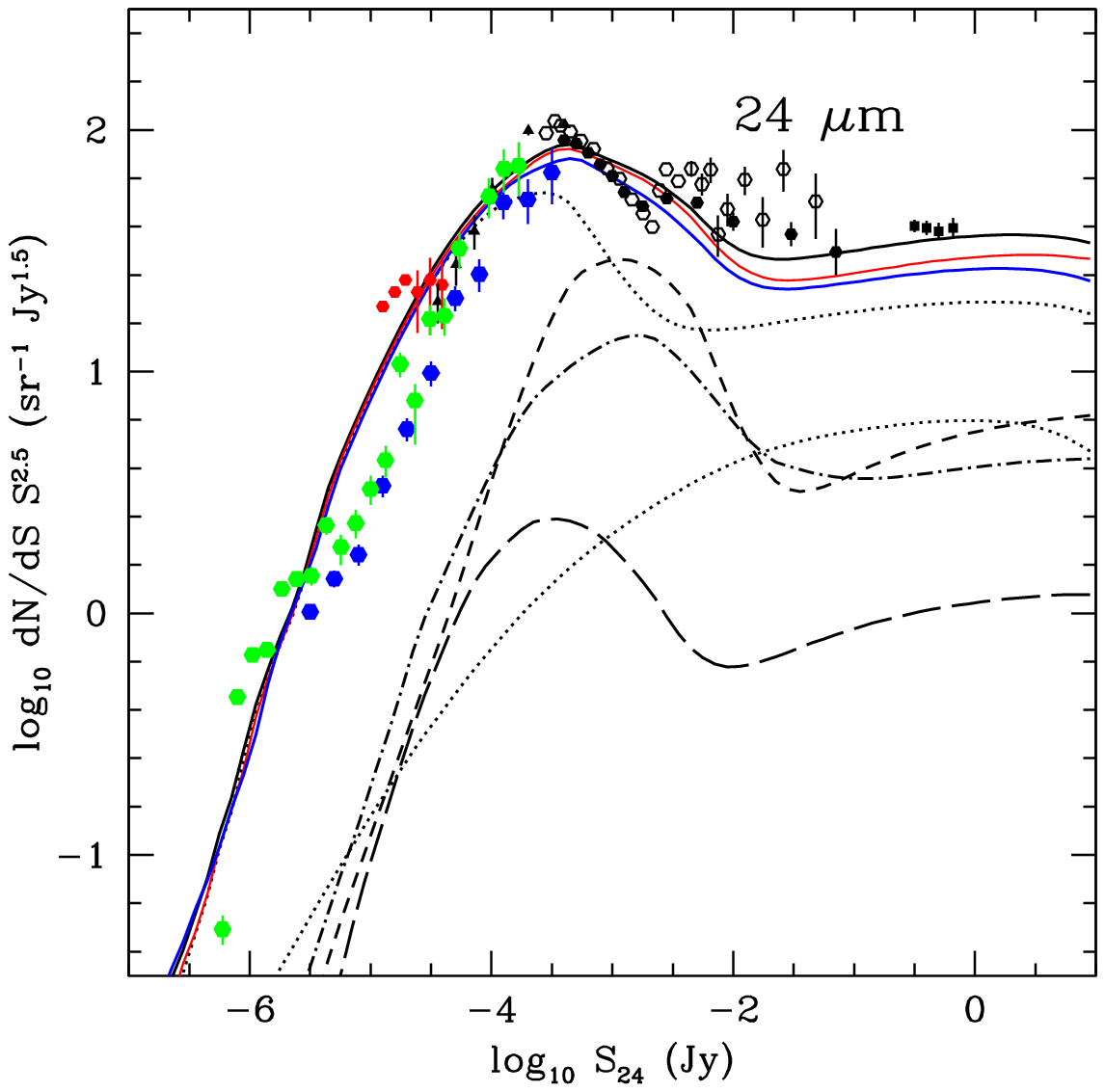,angle=0,width=7.5cm}
\caption{
Euclidean normalised differential counts at {\bf 5.6} (Solid black curve: total counts; dotted curves: post-sb and quiescent (lower); short-dashed curve: M82;
long-dashed curve: A220; and dash-dotted curve: AGN dust tori.  JWST data from Sajkov et al 2024, red filled hexagrams, Stone et al 2024, blue filled
hexagrams, Yang et al 2023, green filled hexagrams, {\it Spitzer} 5.8 $\mu$m data from Davies et al 2021, black filled hexagrams.
JWST 4.5 $\mu$m data from Windhorst et al (2023), blue filled squares, has been included in the 5.6 $\mu$m plot. Predicted counts at 4.5 $\mu$m (magenta locus) 
are indistinguishable from 5.6 $\mu$m counts at faint fluxes.).  Predicted counts from Paper I are shown as a red locus.
{\bf 8} ({\it Spitzer} data from Fazio et al 2004, black filled hexagrams; JWST from Wu et al 2023, 
red filled hexagrams, Stone et al 2024, blue filled hexagrams, Yang et al 2023, green filled hexagrams), 
{\bf 15} ({\it ISO} data from Elbaz et al 1999 (open squares), Aussel et al 1999 (filled squares), Gruppioni et al 2002
(open hexagrams); interpolated {\it IRAS} data (black filled hexagrams) from Verma (private communication); {\it Akari} data from Davidge et al 2015
(filled black triangles); JWST data from Wu et al 2023 (filled red hexagrams), Stone et al 2024, blue filled
hexagrams, Yang et al 2023, green filled hexagrams.
)
and 
{\bf 24}  $\mu$m ({\it IRAS} data from Verma (private communication), filled squares;  {\it Spitzer} data from Shupe et al 2008, 
filled hexagrams, Papovich et al 2004, filled triangles, and Clements et al 2011, black filled squares;
21 $\mu$m JWST data from Wu et al 2023, filled red hexagrams, Stone et al 2024, blue filled
hexagrams, Yang et al 2023, green filled hexagrams   Blue locus is predicted 21 $\mu$m counts.)
}
\end{figure*}


\section{Fits to JWST mid-infrared counts}

Here I discuss the fits to the observed JWST counts at 5.6, 7.7, 15 and 21 $\mu$m, using the models of
Paper I, as modified above.  

Figure 3 show the fits to the counts at 5.6, 7.7, 15 and 24 $\mu$m.  {\it IRAS}, {\it ISO}, {\it Spitzer} and {\it Akari} 
data at 8, 15 and 24/25 $\mu$m are included in these plots.  The Paper I model is shown as red loci and the results
of the improved model in this paper are shown as black loci. To reduce clutter 
on these and subsequent count plots data points are excluded if they are $< 3-\sigma$ measurements.
As noted in Paper I this model provides an excellent fit to the JWST counts at these wavelengths, even 
thought the JWST counts are 100 times deeper than
the {\it Spitzer} counts at 8 and 15 $\mu$m.  The dominant contribution at faint fluxes is from post-starburst 
galaxies, since these dominate at lower infrared luminosities.  Note that at fluxes $<$ 0.1 mJy counts 
at 21 $\mu$m are indistinguishable from those at 24 $\mu$m.
The revised model provides a significant improvement at 5.6 $\mu$m and a small improvement at 7.7 $\mu$m.
The choice of $z_f$ = 10 in Paper I was based on early JWST studies which showed that few galaxies with
z $>$ 10 were being found, but as some galaxies continue to be found with z = 10-16, I also explored the effect 
on the 5.6 $\mu$m counts of changing $z_f$ from 10 to 16.  The predicted counts 
curve was indistinguishable from that for $z_f$ = 10,
so the faint JWST counts give little insight into the evolution of galaxies at very high redshift.

The prediction is that the faint counts at mid-infrared wavelengths are dominated by post-starburst galaxies, reflecting 
the fact that we are mainly seeing galaxies at the faint end of the luminosity function at these flux-densities.

\begin{figure*}
\epsfig{file=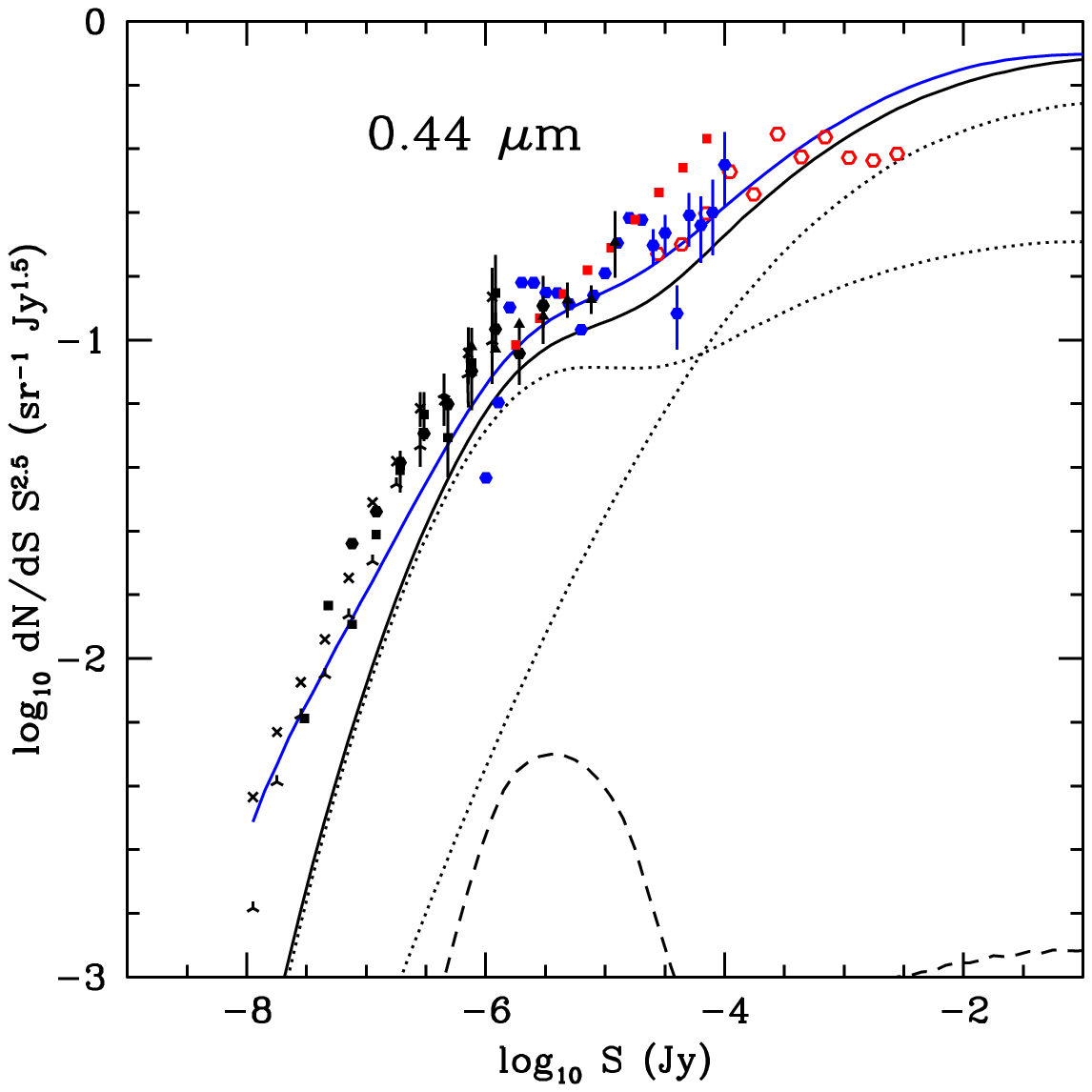,angle=0,width=7.5cm}
\epsfig{file=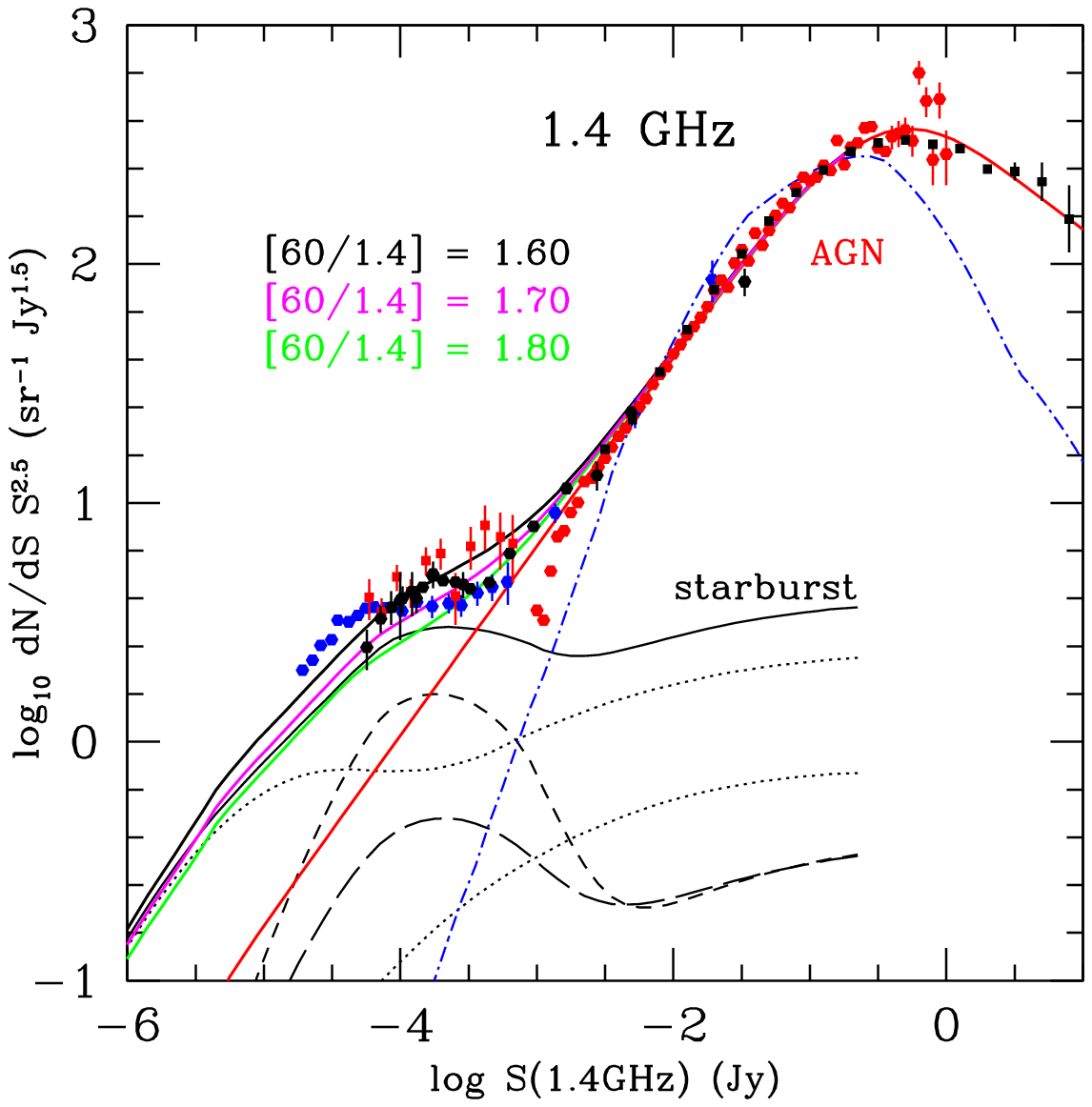,angle=0,width=7.5cm}
\caption{Euclidean normalised differential counts in B (0.44 $\mu$m) (L) and radio (1.4 GHz) (R) bands.
Optical counts data from Koo et al (1986), Maddox et al (1990), Jones et al (1991), Metcalfe et al (1991, 1995, 2001), Stevenson et al (1986).  
The blue curve shows effect of steepened luminosity function at faint luminosities.   Radio counts data from White et al (1997), red
filled hexagrams;  Huynh et al (2005), red filled squares;  Hopkins et al (2003), black filled hexagrams; Matthews et al (2021), 
black filled squares. The deep 3 GHz counts of Smolcic et al (2017) have been included by applying a spectral index of 0.73 (blue filled hexagrams). 
}
\end{figure*}

\begin{figure}
\epsfig{file=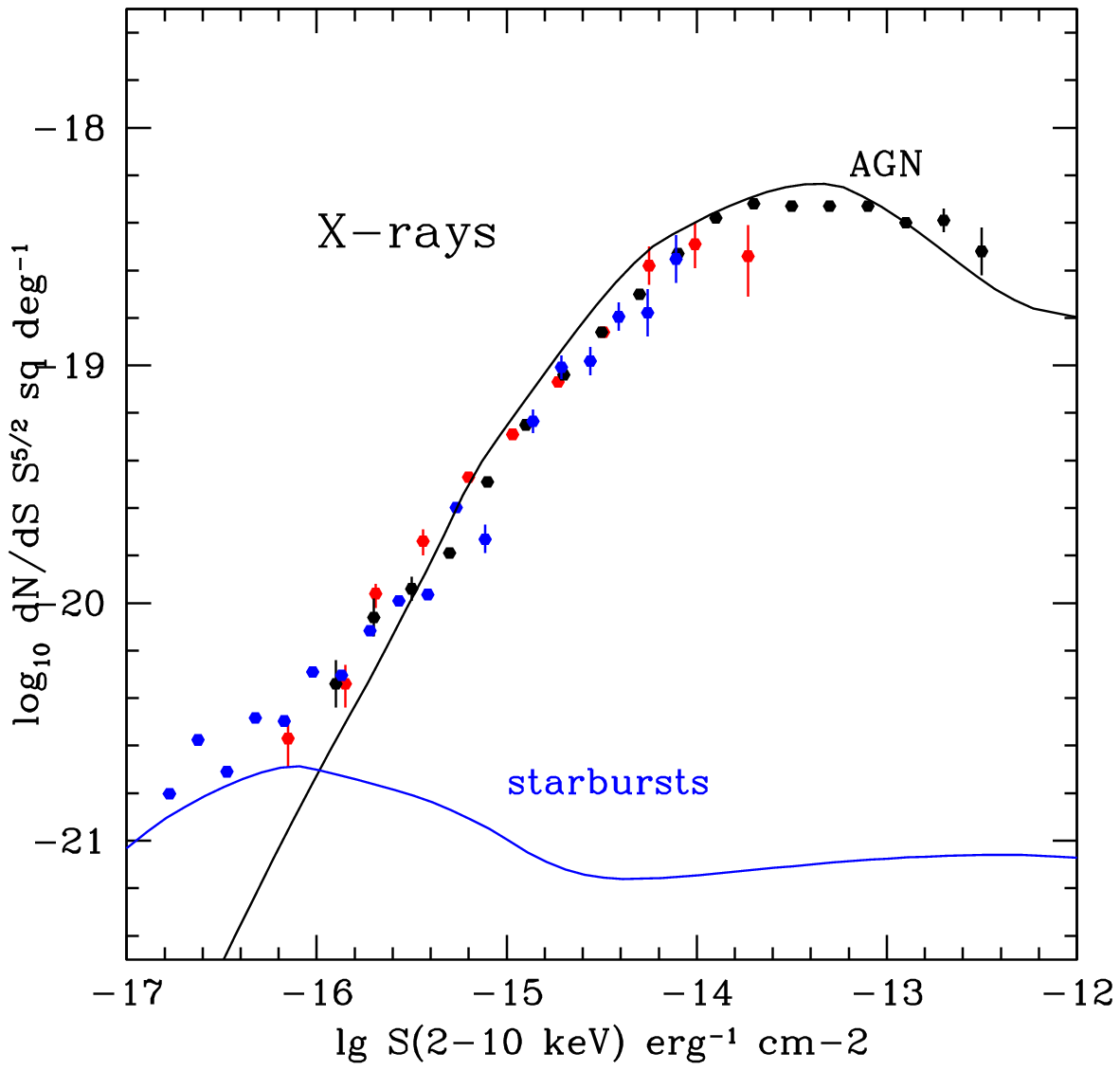,angle=0,width=7.5cm}
\caption{Euclidean normalised differential counts in X-rays (2-10 keV).
Data from Ranalli et al (2003): red hexagons: CDFS data, black hexagrams: XMM-CDFS; Luo et al (2017): blue hexagrams.
Predicted contribution of AGN and starbursts shown separately. 
}
\end{figure}

\section{Extension to optical, radio and X-ray counts}
Figure 4 shows counts in optical (0.44 $\mu$m) and radio (1.4 GHz) bands.  The basic infrared model
underpredicts the faint optical counts, but this can be attributed to the 'faint blue galaxies'.
It is well-known that the faint end of the optical luminosity function steepens at the low luminosity
end as redshift increases (eg Mashian et al 2016).  At later epochs these faint blue galaxies have been swallowed up
in mergers with higher mass galaxies. They do not contribute to the {\it IRAS} 60 $\mu$m luminosity function, so are not 
represented here.  To give an indication of their contribution, the blue locus in Fig 4L corresponds to a steepening 
of the low luminosity slope between $log_{10} L_{60}$ = 7-9 by $\Delta \alpha =0.5$. A more sophisticated treatment 
of this population is beyond the scope of this paper. 

To predict counts at radio wavelengths, I use the approach of Rowan-Robinson et al (1993).  For radio-loud
galaxies due to massive black holes in elliptical galaxies and QSOs I use the luminosity function and
empirical pure luminosity evolution rate of Dunlop and Peacock (1990), with a small adjustment to allow for the different
 cosmological model used here.  For radio-quiet galaxies with radio
emission due to star-formation in spiral galaxies, I use an assumed ratio of $L(60)/L(1.4 GHz)$, based
on the analysis of De Zotti et al (2024), after applying suitable bolometric corrections for each 
(non-AGN) component.  The De Zotti et al $z \sim 2 - 3.5$ values for $q_{IR}$, 2.20,  translate to $log_{10} L(60)/L(1.4 GHz)$
of 1.52, 1.85 and 1.89 for cirrus, M82 and A20 templates.  Figure 4R shows predicted counts for 
$log_{10} L(60)/L(1.4 GHz)$ = 1.6, 1.7 and 1.8.  The dash-dotted locus shows an attempt to derive the AGN counts from
the infrared dust torus component.  Although the latter traces the black hole optical and uv luminosity well, the link
to radio emission is obviously more complex, involving the magnetic field and radio jet structure.

X-ray source-counts can also be modelled, using the relationship between AGN dust torus and
X-ray luminosity discussed by Rowan-Robinson et al (2009), ie assuming
\newline

$lg L_{tor} \sim lg L_{opt}$ - 0.4     (3) (assuming a dust torus covering factor of 40$\%$),
\newline
where $L_{tor}$ is the 1-1000 $\mu$m bolometric luminosity in the dust torus component,
\newline

$lg L_X \sim lg L_{opt}$ + 0.3      (4)   (far uv bolometric correction),
\newline
where $L_{opt}$ is the 0.1-3 $\mu$m bolometric luminosity,
\newline

$lg L_X = lg L(2-10 keV)$ + 1.43    (5)  (hard X-ray bolometric correction),
\newline
where $L_X$ is the 0.1-100 keV bolometric luminosity,
\newline

and $lg L_{tor} = lg(\nu(24 \mu m) L(24 \mu m))$ + 0.73  (6)  (dust torus bolometric correction at 24 $\mu$m).
\newline

Neglecting the small differences in K-corrections, these combine to give
\newline

\noindent
$S(2-10 keV,erg s^{-1})_{AGN} \approx 10^{-10.27} k_1 S(24\mu,Jy)_{tor}$  (7)
\newline

where $k_1$ is an adjustable factor allowing for the approximate nature of equations (3) and (4).

The contribution of starbursts to the X-ray counts can be modelled using the relation between X-ray and far infrared
luminosities give by Ranalli et al (2003).
\newline

$lg L(2-10 keV) \sim lg L_{fir}$ - 3.68.   (8)
\newline

Neglecting the small differences in K-corrections, this gives
\newline

$S(2-10 keV, erg s^{-1})_{sb} \approx 10^{-12.99} k_2 S(24 \mu, Jy)_{sb}$    (9)
\newline

where $k_2$ is an adjustable factor allowing for the approximate nature of equation (8).

Figure 5 shows the predicted X-ray counts for AGN and for starbursts compared with the 2-10 keV
counts of Ranalli et al (2013), with $k_1 = 0.49, k_2 = 1.8$ (chosen to give the best fit to the X-ray counts). Interestingly, 
starbursts start to dominate the counts at the faint end, as in the radio case.

\begin{figure*}
\epsfig{file=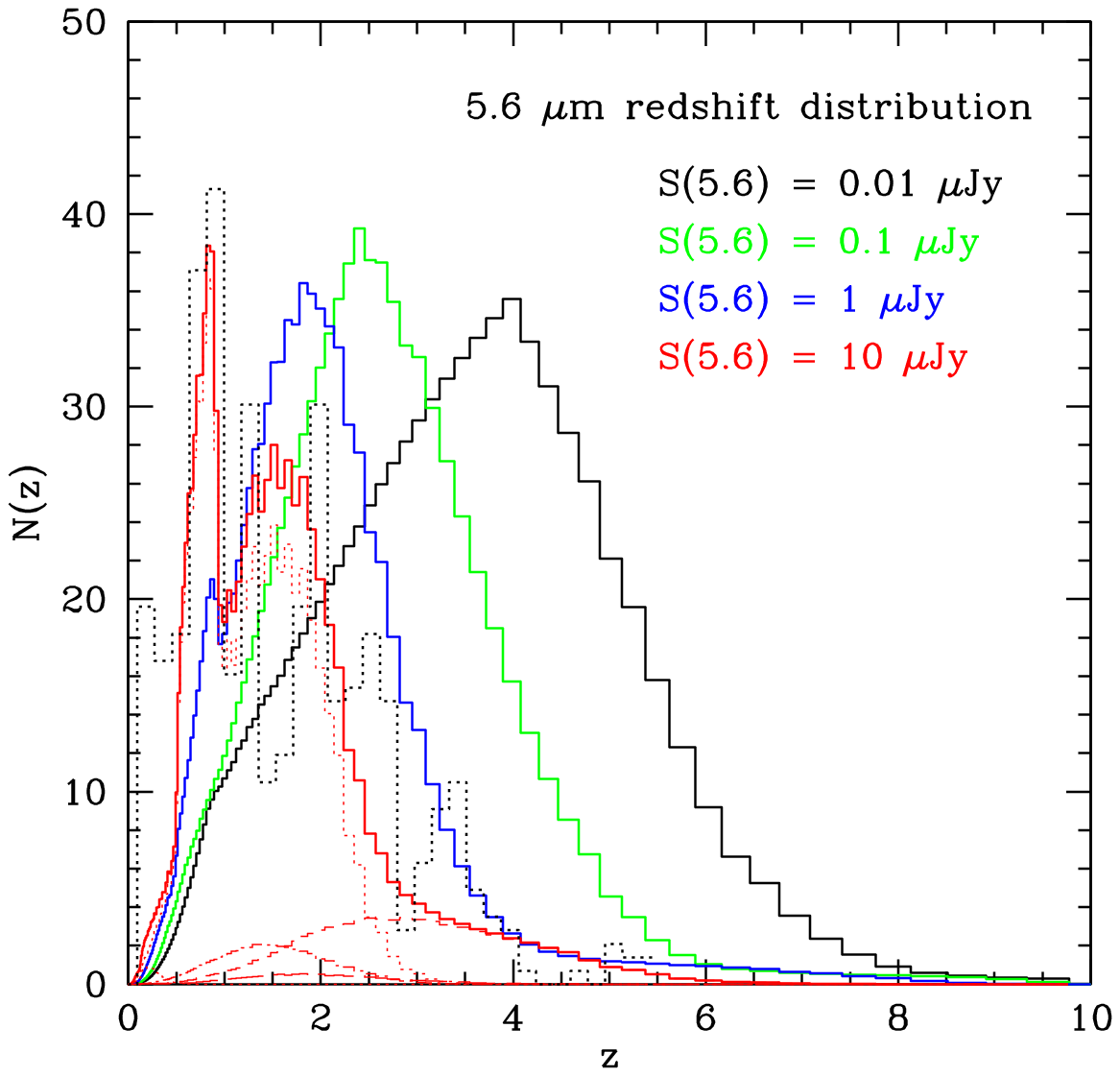,angle=0,width=14cm}
\caption{Predicted 5.6 $\mu$m normalised redshift distribution at S(5.6) $>$ 10, 1, 0.1, 0.01 $\mu$Jy. For the brightest bin 
the distributions are show separately for post-starburst and quiescent (dotted), interaction (dashed) and merger (long-dashed), and
AGN dust torus (dash-dotted) populations.  The observed redshift distribution of Sajkov et al (2024) is shown as a dotted black histogram.
}
\end{figure*}

\begin{figure*}
\epsfig{file=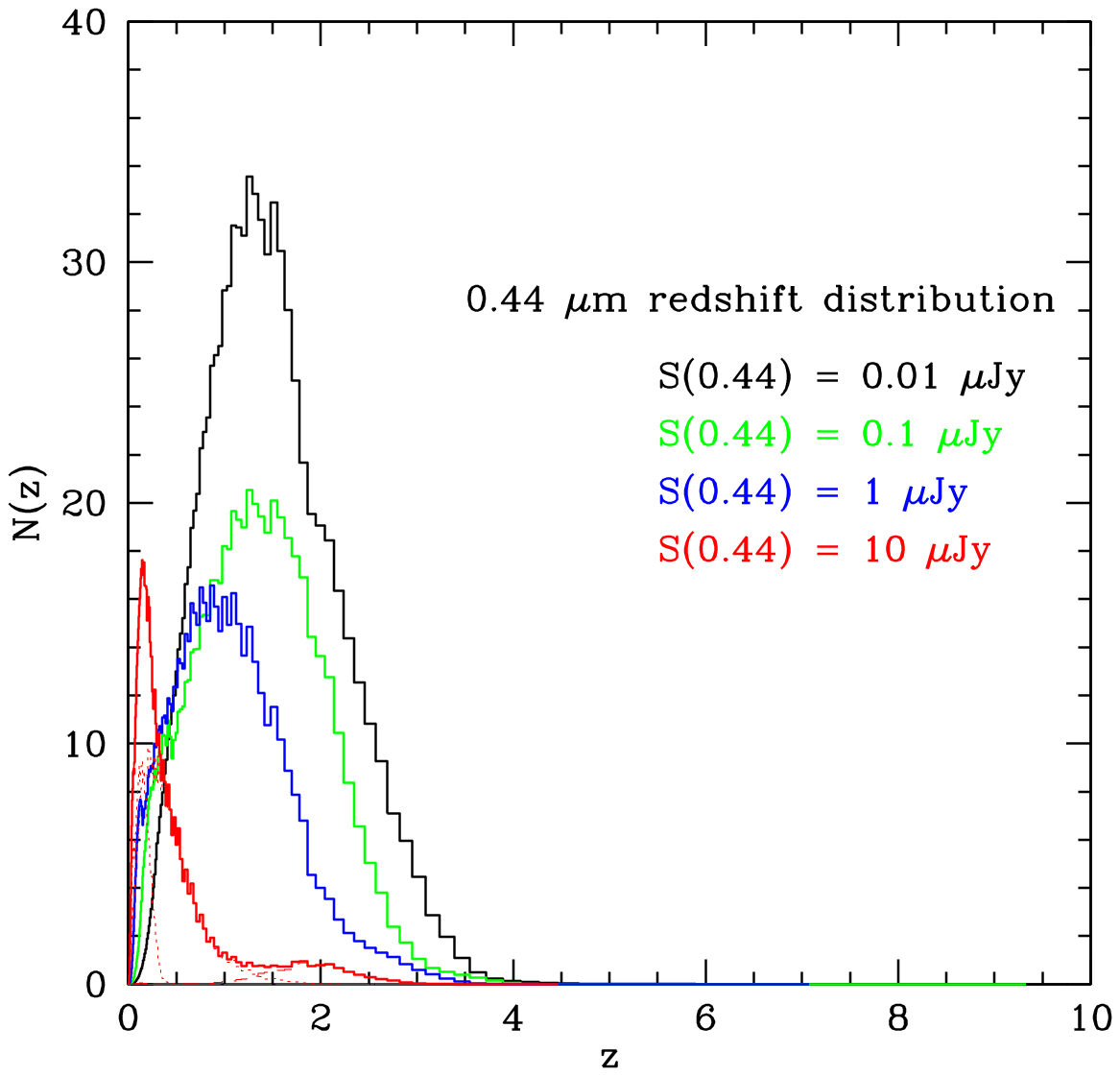,angle=0,width=7.4cm}
\epsfig{file=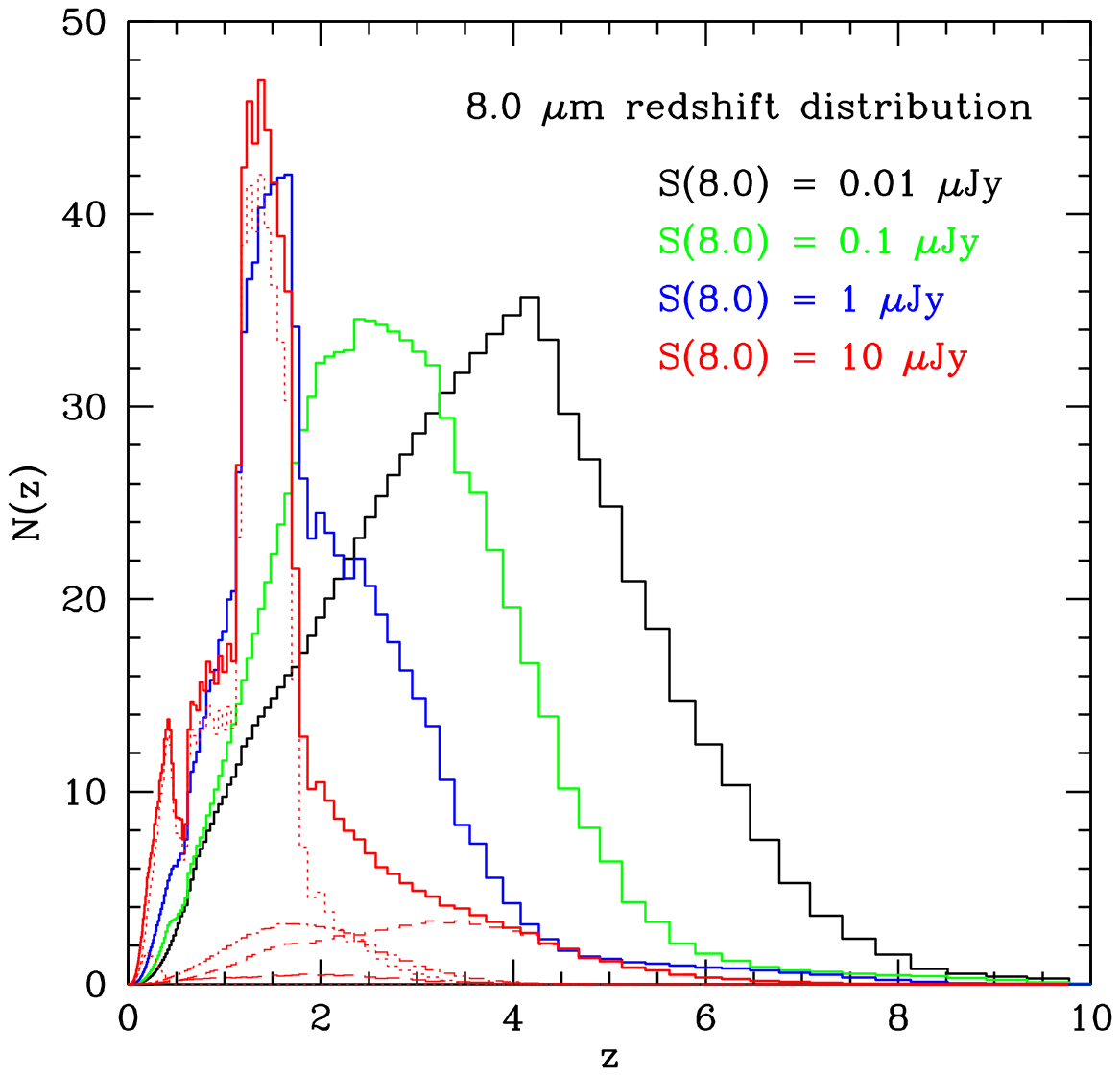,angle=0,width=7.4cm}
\epsfig{file=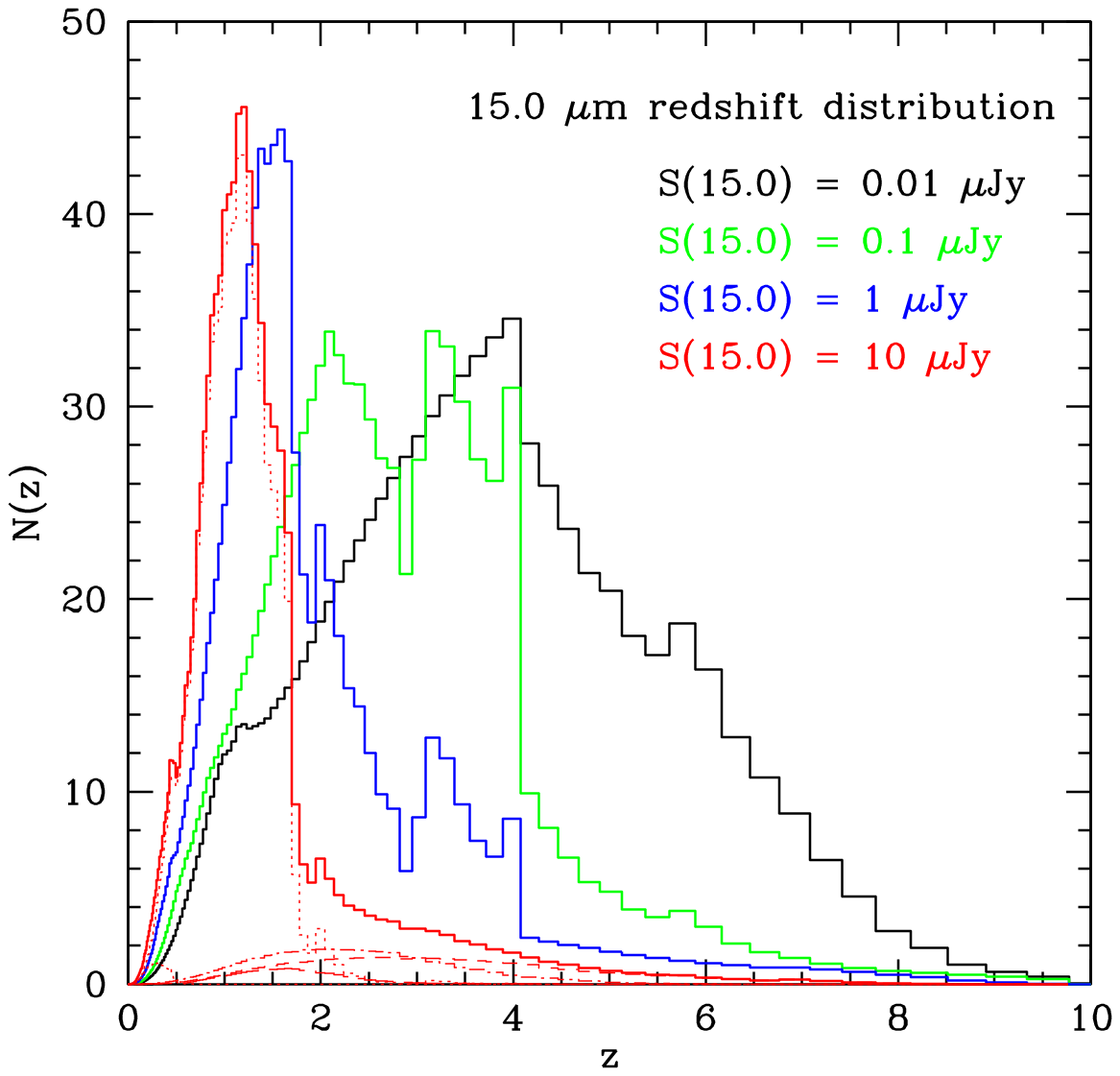,angle=0,width=7.4cm}
\epsfig{file=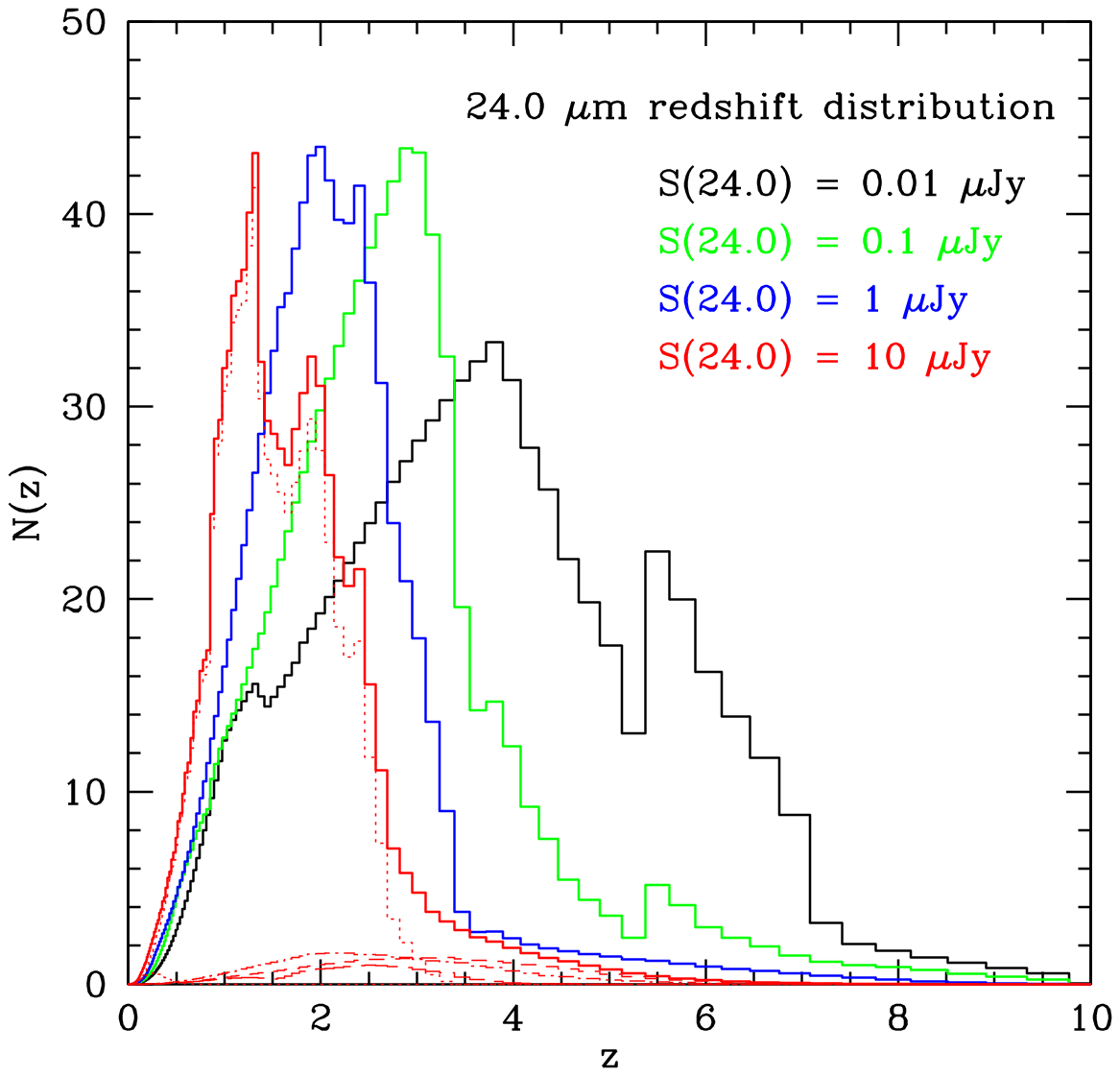,angle=0,width=7.4cm}
\epsfig{file=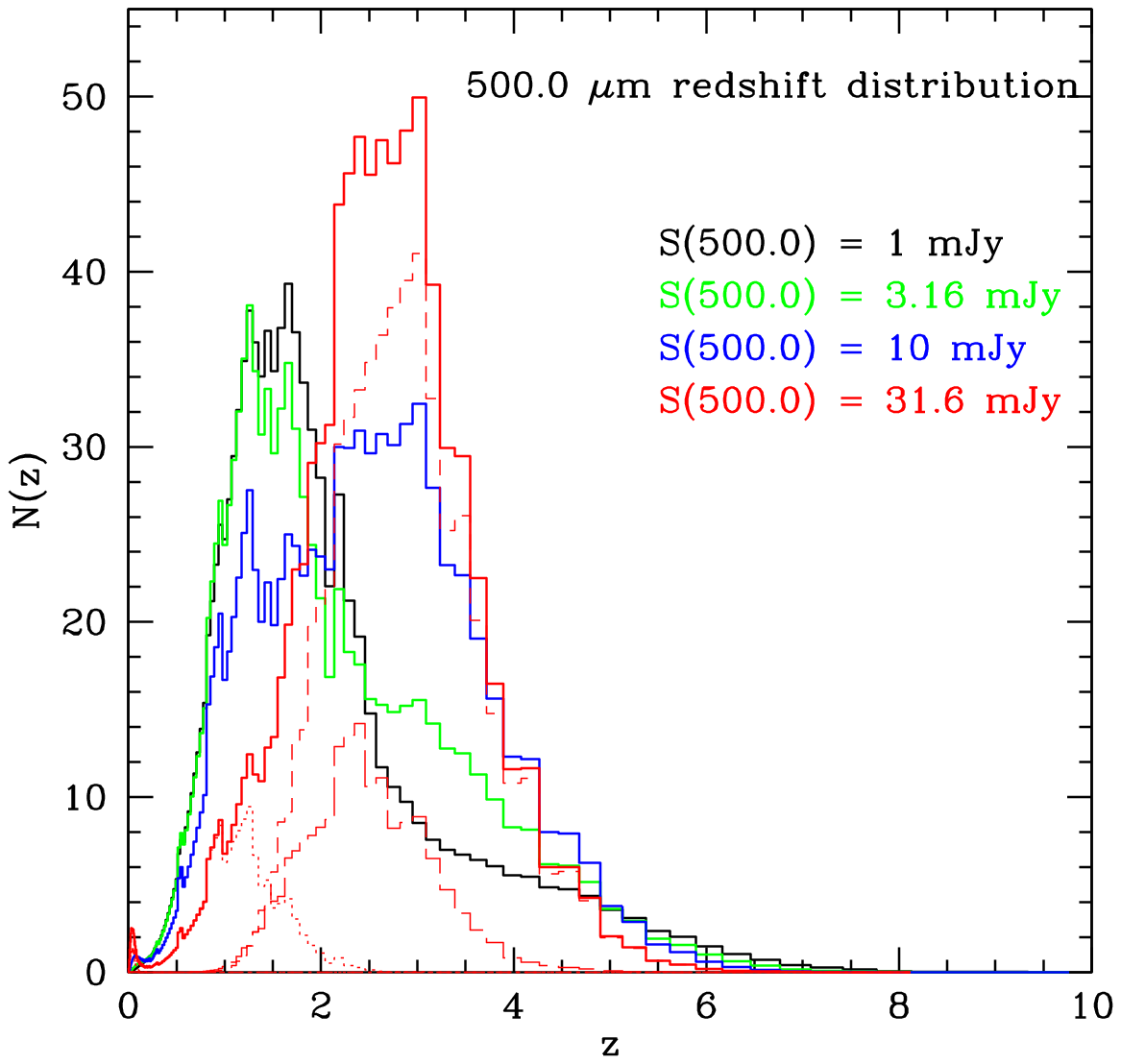,angle=0,width=7.4cm}
\epsfig{file=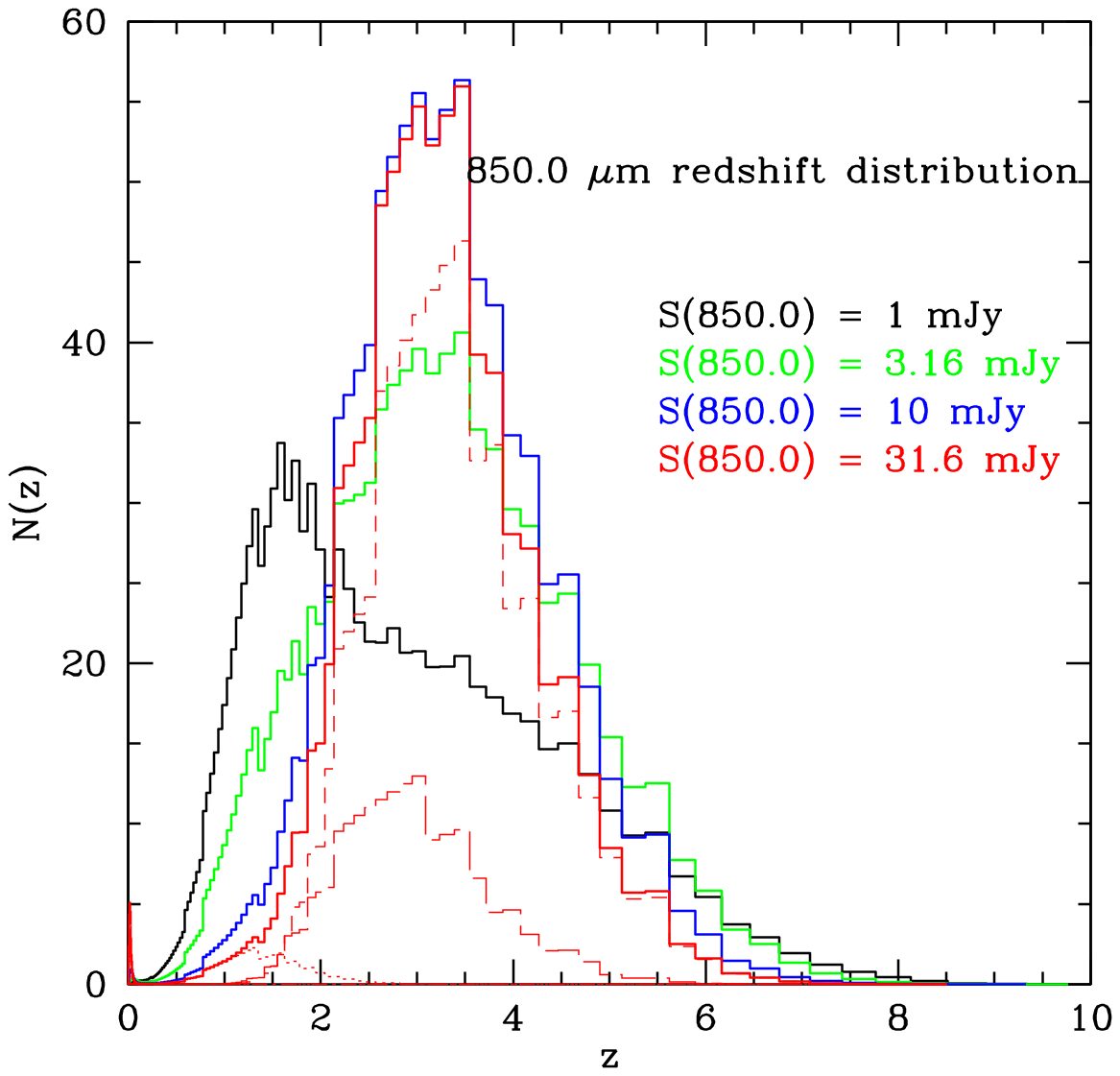,angle=0,width=7.4cm}
\caption{Normalised redshift distribution at 0.44, 8, 15, 24, 500 and 850 $\mu$m. As in Fig 6, for the brightest bin 
the distributions are show separately for the different populations.
}
\end{figure*}

\section{Redshift distributions}
The models also yield redshift distributions at each wavelength and flux-level and a selection of 
these are illustrated in Figs 6 and 7.  In Fig 6 I have included the 5.6 $\mu$m redshift distribution given by Sajkov et al (2024),
shown as a black dotted histogram, which appears to be intermediate between the predictions for S(5.6 $>$ 10 $\mu$Jy and
1 $\mu$mJy.  The 5.6 $\mu$m distributions shown in Fig 6 are broadly consistent with
the flux-z distribution shown by Ostlin et al (2025) in their Fig 14.  The model predicts a long tail of starburst 
galaxies with z = 2-6 for S $>$ 10 $\mu$Jy, caused by the negative K-correction at
5.6 $\mu$, combined with the strong luminosity evolution.  As mentioned in section 2 an adjustment was made to the amplitude of the 
optical component of the M82 starburst to prevent an obvious conflict with the observed redshift distribution of Ostlin et al. 

At 500 and 850 $\mu$m (see Fig 7), the
strong negative K-correction also predicts the curious effect that at bright fluxes, the fraction of sources at higher
redshift (2-5) is much greater than is predicted at fainter fluxes.  The combination of strong 
luminosity evolution in the starburst galaxy types with the strong negative K-correction yields
a secondary peak at high redshifts, which is dominant at bright submillimetre fluxes.
The strong sensitivity of these redshift distributions to PAH and other spectral features demonstrates
the potential of the JWST photometry for photometric redshift estimation (eg Rieke et al 2024).

At 0.44 $\mu$m redshifts are limited to z $<$ 4 by the Lyman cutoff.

\section{Integrated background radiation}

Earlier models of the integrated background spectrum were discussed in Paper I.


Figure 8 shows the predicted
background spectrum derived from our revised counts model.  The fit from mid infrared to submillimetre wavelengths is excellent.  
We also show the fit in the optical and near IR, which reflects the assumed optical templates. 
The fit at these wavelengths is surprisingly good, considering the simplified nature of the modelling at these wavelengths.
The main contribution at all wavelengths comes from post-starburst galaxies.

\begin{figure}
\epsfig{file=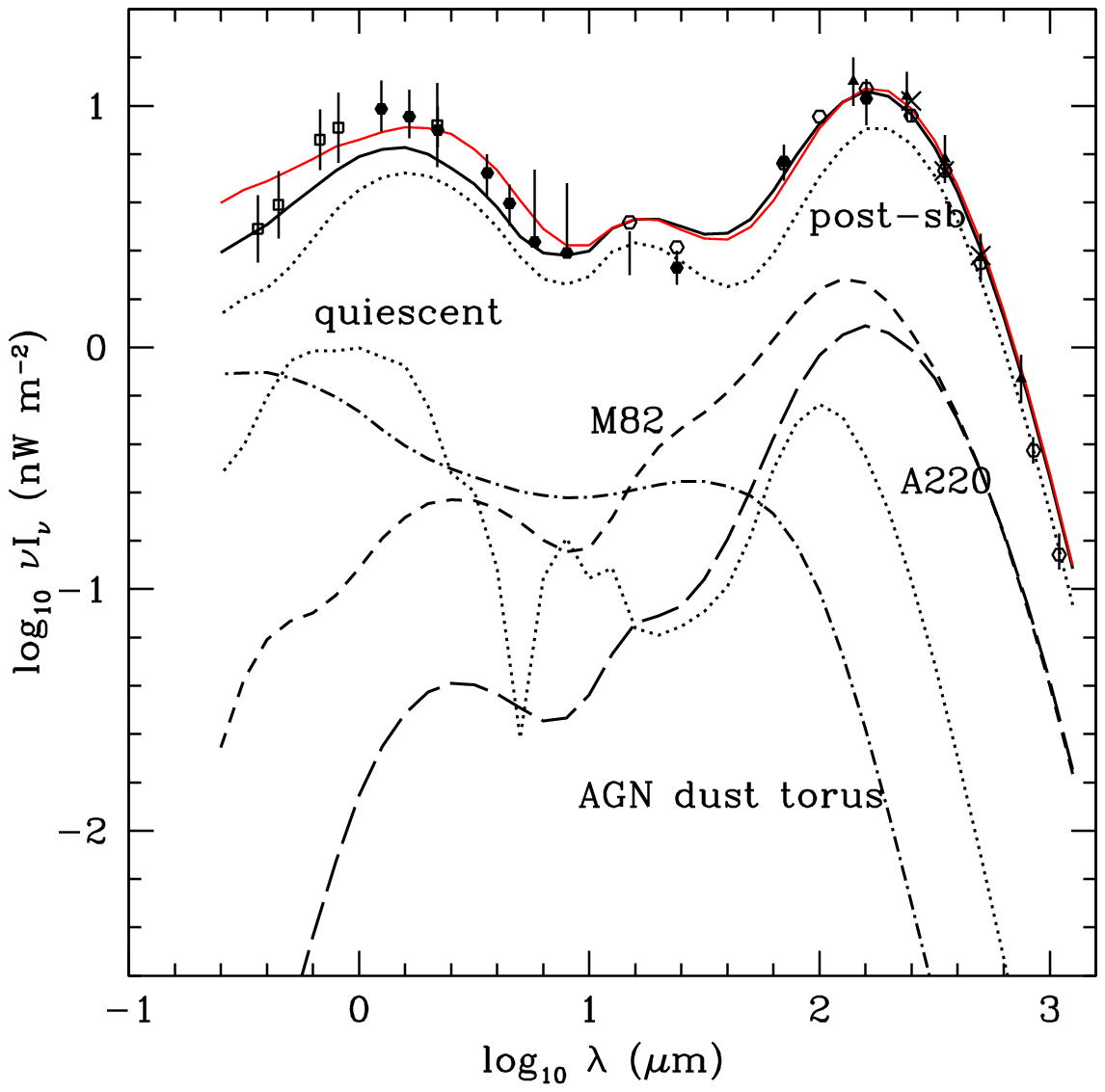,angle=0,width=7.5cm}
\caption{
Background spectrum from 0.25 to 1250 $\mu$m. Data from Fixsen et al (1998, open triangles),
Dole et al (2006, filled circles), Serjeant et al (2000, vertical bar), Pozzetti et al (1998, open squares), Kashlinsky (2005), 
Bethermin et al (2012, open circles).
Solid black locus: predicted background for model with SEDs and evolution as in Paper I except for revised treatment of post-sb and 
quiescent components. Other loci: as in Fig 3.
}
\end{figure}

%
%
%
%

\section{Discussion}

The model presented here, based on five basic infrared templates is inevitably an oversimplification.  
We know that to understand local galaxies we need a range of cirrus components (eg Rowan-Robinson 1992, 
Efstathiou and Rowan-Robinson 2003).  
From the models of Efstathiou et al (2000) the SED of a starburst varies strongly with the age of the starburst
so the representation as three simple extremes, M82 and Arp220 starbursts, and post-starbursts, is oversimplified.
Similarly a wide range of AGN dust torus models are needed to characterise their observed spectra (eg Efstathiou and
Rowan-Robinson 1995, Farrah et al 2003, Rowan-Robinson and Efstathiou 2009, Siebenmorgen et al 2015).

However the five templates used here do capture the main features of the infrared galaxy population.  In fitting
the SEDs of {\it ISO}, {\it Spitzer} and {\it Herschel} sources (Rowan-Robinson et al 2004, 2005, 2008, 2014), we allowed individual sources
to include cirrus, M82 or Arp220 starburst, and an AGN dust torus, which gives a rich range of predicted
SEDs.  This works for SCUBA sources (Clements et al 2008) and for detailed IRS spectroscopy of
infrared galaxies (Farrah et al 2008).

From Fig 2 the interactions and mergers starbursts peak at t = 2 Gyr, with post-starburst galaxies lagging about 1 Gyr after this.
Quiescent galaxies become dominant in the past few billion years of the universe's history.
AGN dust tori, where the evolution reflects the black hole accretion history, also peak at t = 2 Gyr, but show a less steep
decline to the present than the
starbursts, suggesting a gradual change in the relative efficiency with which gas is converted to stars
or accreted into a central black hole (cf Rowan-Robinson et al 2018). 

An interpretation of our assumed form of evolution is as follows: the power-law term represents the rapid increase of 
star-formation in the universe during the merger and interaction phase of galaxy assembly, while the exponential decline 
corresponds to the gradual exhaustion of gas in galaxies by star-formation.  Rowan-Robinson (2003) showed how 
this form of evolution fits into a simple closed box model for star-formation in galaxies.

An accompanying data file gives predicted integral counts at selected wavelengths from 5.6 to 24 $\mu$m.

This backward evolution approach could be improved with a better understanding of dust evolution and of 
the evolution of the luminosity function at high redshift.  A very different, and in principle more illuminating, approach is followed by Cowley et al
(2018), in which the GALFORM galaxy formation simulation is combined with a dust model to predict source-counts 
in the JWST mid-infrared wave-bands, with some success at faint fluxes (Wu et al 2023).

An appendix discusses the analogy between the star-formation history in galaxies with the outbreak of an epidemic like
COVID-19. 

\section{Conclusions}

The counts model of Rowan-Robinson (2009), as modified in Paper I and here, provides an excellent fit to the deep JWST counts at 5.6-21 $\mu$m, and to the integrated background spectrum.  There is almost no change to the fits in Paper I to far infrared and submillimetre counts.  The models have also been extended to the optical B-band and to radio and X-ray wavelengths.

The last sixty years have seen the opening up of the radio, X-ray, millimetre-wave, mid and far infrared and submillimetre bands, in which the roles of first AGN and then 
starburst galaxies have been revealed.  It is satisfying to pull the surveys at these wavelengths together into a unified evolutionary picture.


\section{Acknowledgements}
I thank Meredith Stone for helpful comments on an earlier version of this work. I thank Duncan Farrah and Lingyu Wang for helpful comments.
An anonymous referee made helpful comments which have resulted in improvements to the paper.

\section{Data availability}
All data used in this paper are in the public domain.


\section{Analogy with spread of an epidemic}

The form of evolution used here for star-formation in galaxies, consisting of a rapid rise followed by an exponential decay,
can also be applied to the outbreak of an epidemic like that of COVID-19 in 2020. Figure 9 shows the number of cases each day,
and the number of deaths, in Italy and the UK for the first 60 days of the outbreak, modelled with a $t^P exp -Qt$ function.
Here the exponential decline is due to the gradual depletion of the infectable population.

\begin{figure}
\epsfig{file=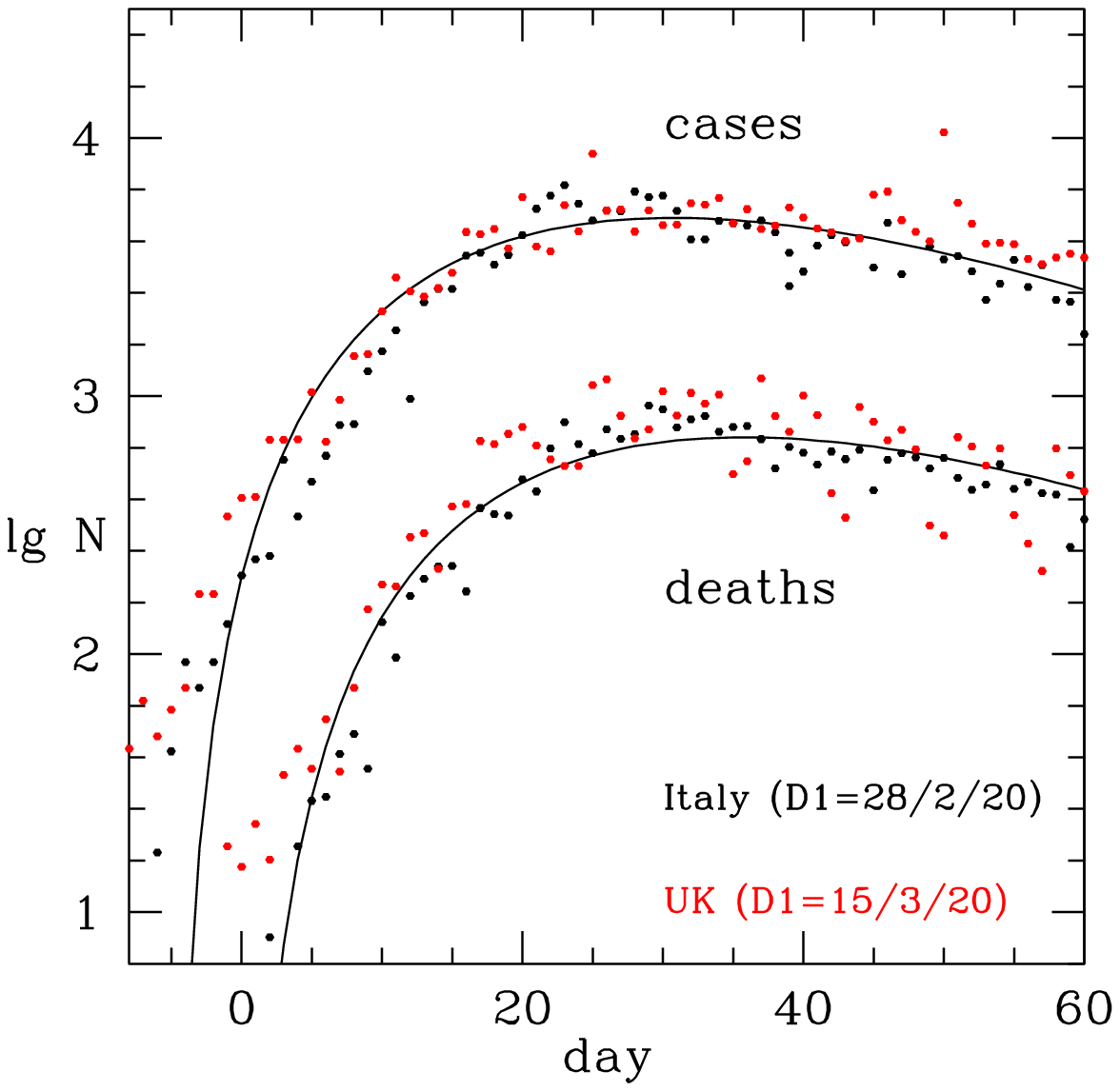,angle=0,width=7.5cm}
\caption{The deaths per day, and the number of cases, of COVID-19 in Italy and the UK during
the first 60 days of the 2020 outbreak, modelled with a simple $(t/t_0)^P exp -Qt/t_0$ form, with P = 2.9, Q = 8.1, $t_0$ = 100 days  On average, at this stage, 1.4$\%$ of cases resulted in death 7 days later.
}
\end{figure}

\end{document}